\documentstyle[aps,floats,preprint,rotating,epsfig]{revtex}
\tightenlines

\begin{document}
\draft
\preprint{\vbox{
\hbox{hep-ph/0108258}
\hbox{July 2001}}}
\title{Signatures of Doubly Charged Higgs Bosons in $e\gamma$ 
Collisions}
\author{Stephen Godfrey, Pat Kalyniak and Nikolai Romanenko}
\address{
Ottawa-Carleton Institute for Physics \\
Department of Physics, Carleton University, Ottawa, Canada K1S 5B6}

\maketitle

\begin{abstract}
We study the discovery potential for doubly charged Higgs bosons, 
$\Delta^{--}$, in 
the process $e^-\gamma\to e^+ \mu^-\mu^-$ for centre of mass energies 
appropriate to high energy $e^+e^-$ 
linear colliders and the CLIC proposal.
For $M_\Delta < \sqrt{s_{e\gamma}}$ discovery is likely for even
relatively  small values of the Yukawa coupling to leptons.  However, 
even far above threshold, 
evidence for the $\Delta$ can be seen due to contributions from 
virtual intermediate $\Delta$'s although, in this case,
$\mu^-\mu^-$ 
final states can only be produced in sufficient numbers for discovery 
for relatively large values of the Yukawa couplings.
\end{abstract}

\section{Introduction}

Doubly charged Higgs bosons arise in many extensions of the standard model 
(SM), typically as components of $SU(2)_L$ triplet representations.  
Although the single Higgs doublet adopted to break the SM gauge symmetry 
is  the simplest possibility for the Higgs sector, the real nature of the
Higgs sector is unknown and many other cases are worth consideration.
A simple extension, which arises naturally in 
Supersymmetry, is to include two Higgs doublets.  Beyond this, the 
introduction of a Higgs triplet is one of the next logical 
possibilities for the Higgs sector.   While in the context of the SM
there is no specific 
motivation for the introduction of a Higgs triplet
there are several models which require a Higgs triplet for 
symmetry breaking.  Perhaps the 
best known model with this requirement is the Left-Right symmetric 
model \cite{LR}.  
In this model the neutral scalar couplings to the fermions may also 
give rise to the 
see-saw mechanism leading to naturally small neutrino masses \cite{seesaw}.
Another example of a model containing a Higgs triplet is 
the Left-handed Higgs triplet model of 
Gelmini and Roncadelli \cite{Gelmini}.
One of the consequences of a Higgs triplet with the appropriate 
quantum numbers 
is the existence of a
doubly charged Higgs boson, the $\Delta^{--}$, which has a 
distinct experimental signature.  
Although the introduction of a Higgs triplet can introduce 
phenomenological difficulties it turns out that they are not difficult 
to avoid.
Thus, the discovery of a $\Delta^{--}$ would have important 
implications for our understanding of the Higgs sector and more 
importantly, for what lies beyond the standard model.

There are a variety of processes sensitive to doubly charged Higgs 
bosons.
Indirect
constraints on masses and couplings have been obtained from lepton
number 
violating processes and muonium-antimuonium conversion experiments 
\cite{swartz,hm,fujii,chk}.  The most stringent limits come from that latter 
measurement \cite{muoniumex,swartz}.  For flavour diagonal couplings these measurements 
require that the ratio of the Yukawa coupling, $h$, and Higgs mass, 
$M_\Delta$, satisfy
$h/M_\Delta < 0.44$~TeV$^{-1}$ at 90\% C.L..
These bounds allow the existence of low-mass doubly charged Higgs with 
a small coupling constant.  These limits 
can be 
circumvented in certain models as a result of cancellations 
among additional diagrams arising from other new physics 
\cite{framras,pleitez}.  Thus, 
from this point of view, direct limits are generally more robust.

Search strategies for the $\Delta^{--}$ have been explored for the 
Fermilab Tevatron and the CERN 
Large Hadron Collider (LHC) \cite{datta,gunion}. At the Tevatron it is 
expected that the doubly charged Higgs can be detected
via pair production if its mass is 
less than $\sim 275$~GeV while at the LHC the reach extends to 
$\sim 850$~GeV, in 
both cases assuming a BR to leptons of 100\% \cite{datta}.
Signatures for the
 $\Delta^{--}$ have also been explored for high energy $e^+e^-$ 
colliders \cite{cuypers}.  For the most part, limits obtained at $e^+e^-$ 
colliders have relied on $\Delta^{--}$ pair production so the
mass reach is limited to $M_\Delta < \sqrt{s}/2$ although these can be 
exceeded by constraining   t-channel $\Delta$ exchange in Bhabba 
scattering and single $\Delta^{--}$ production.   
Studies looking at doubly charged 
Higgs production in $e^-e^-$ colliders \cite{framras,cuypers,emem}
find mass limits up to $\sqrt{s}$. 
There have been a number of studies of single $\Delta^{--}$ in 
$e\gamma$ \cite{cuypers,london,gregores,rizzo}
collisions and also $e^+e^-$ and $\gamma\gamma$ collisions 
where in the latter cases the photon or electron is described using 
the effective photon or effective fermion approximation respectively 
\cite{london}.
In the case of $e\gamma$ collisions  the kinematic limit is $\sim 
\sqrt{s_{e\gamma}}$.  
The most recent calculation by Gregores {\it et al}. \cite{gregores}, 
which most closely resembles the approach presented here,  
only included the
Feynman diagrams with s-channel contributions from the $\Delta^{--}$
and therefore restricted their study to resonance $\Delta^{--}$ production.
In effect they looked at $e^-e^-$ fusion where one of the $e^-$'s is 
the beam electron and the other arises from the equivalent 
particle approximation for an electron in the photon.  In this 
approximation the authors assume that the positron is lost down the 
beam.

In this paper we study signals for doubly charged Higgs bosons in the 
process $e^-\gamma\to e^+ \mu^-\mu^-$ including all contributions to 
the $e^+\mu^-\mu^-$ final state.  
We assume the photon is produced by backscattering a laser from the
$e^+$ beam of an $e^+e^-$ collider \cite{backlaser}.
We consider $e^+e^-$ centre of mass 
energies of $\sqrt{s}=500$, 800, 1000, and 1500~GeV
appropriate to the TESLA/NLC/JLC high energy colliders 
\cite{tesla,nlc,jlc}
and $\sqrt{s}=3$, 5, and 8~TeV appropriate to the CLIC proposal 
\cite{clic}.  In 
all cases we assume an integrated 
luminosity of ${\cal L}=500$~fb$^{-1}$.
Our calculation includes diagrams which would not contribute to on-shell 
production of $\Delta^{--}$'s.  Because the signature of same sign 
muon pairs in the final state is so distinctive and has no SM background, 
we find that the 
process can be sensitive to virtual $\Delta^{--}$'s with masses in
excess of
the centre of mass energy, depending on the strength of the
Yukawa coupling to leptons.  

In the next section we give a short overview of some models with 
doubly charged Higgs bosons and write down the Lagrangian describing 
their couplings to leptons. We also note some of
the existing constraints on the relevant parameters of these models.  
In section III we describe our 
calculations and present our results.  We conclude in section IV with 
some final comments.

\section{The models}

In this section we give a brief description
  of the two most popular models with triplet Higgs bosons: 
The Left-handed Higgs Triplet Model
(LHTM) and the Left-Right symmetric model (LRM). We mention only those
aspects of the models which are of particular relevance to the process
under consideration here. This, of course, includes the form of the Yukawa 
coupling
of the triplet Higgs boson to fermions and we mention existing limits. 
We also consider
the size of the possible vacuum expectation value of the neutral
component of the triplet and include some discussion of the scalar mass 
spectrum of the models. All these parameters dictate which decay modes
exist for the doubly charged Higgs and, consequently, its width. The
process we consider is sensitive to the width of the $\Delta$
when $M_\Delta < \sqrt{s_{e\gamma}}$.
All our results are applicable to both models.

\medskip

{\it The Left-handed Higgs Triplet model (LHTM)}
   contains at least one Higgs triplet with weak 
   hypercharge $Y=2$ which has lepton number violating couplings
    to leptons but  does not couple to quarks. 
This triplet Higgs field was first introduced
by Gelmini and Roncadelli \cite{Gelmini}
in order to give rise to Majorana masses for left-handed neutrinos 
while
preserving $SU(2)_L$ gauge symmetry. The LHTM contains a
Higgs triplet field in addition to all SM matter particles
and the usual Higgs doublet.
The minimal Higgs multiplet content of the model is thus \cite{Godbole}
\begin{equation}
\Delta=
\left( \begin{array}{cc}
\Delta^-/\sqrt{2} & \Delta^{--} \\
\Delta^0 & -\Delta^-/\sqrt{2} 
\end{array} \right), \,\,\, \phi=
\left( \begin{array}{c}
\phi^+ \\
\phi^0 \end{array} \right).
\end{equation}
Here the neutral components of the triplet and doublet multiplets may
get vacuum expectation values (VEVs), denoted as 
$\langle\Delta^0\rangle = w/\sqrt{2}$ and
$\langle\phi^0\rangle =v/\sqrt{2}$, respectively.

Isotriplet contributions to the masses of the electroweak bosons would 
result in a  $\rho$-parameter, 
$\rho = (1+2x^2)/(1+4x^2)$ where $x = w/v$,
in the LHTM which is less than unity even
at tree level. As a result, the VEV, $w$, of the triplet Higgs 
is constrained to be small compared with the VEV, $v$, of the doublet.
It is natural to be guided by the LEP II bound on the $\rho$ parameter 
\cite{LEP}. The 3$\sigma$ LEP II bound corresponds to 
$w \sim 15$ GeV.
Thus, we expect the value of the triplet's VEV
to be less than this and it may reasonably be set to zero; 
this is further justified below based on neutrino mass arguments.

The physical scalar particle content of the model is as follows:
doubly charged Higgs $\Delta^{--}$, 
singly charged Higgs $H^-$
 which are a mixture of the triplet and
doublet components,
and three neutral
particles (two scalars and one pseudoscalar).
The mixing in the neutral scalar sector is governed by 
the scalar potential, the most general form of which may be found in
\cite{Godbole}.

The Higgs potential of ref. \cite{Godbole} yields the 
following approximate relation between 
various scalar masses, valid in the limit $w \rightarrow 0$:
\begin{equation}
 M_{\Delta^{--}}^2+ M^2_{A}=2 \cdot M_{H^-}^2
 \label{massrelation}  
\end{equation}
where $M_{A}$ is the pseudoscalar (pseudomajoron) mass. This 
relation implies
that masses of doubly and singly charged Higgs particles should not 
differ too much
(for reasonable Higgs self couplings). This has implications regarding the 
decay modes of the $\Delta^{--}$.

The triplet's Yukawa coupling to lepton doublets  is given by 
\begin{equation} 
\label{eq1}
{\cal L}_{Yuk}=-i h_{ll'} \Psi_{l L}^T C \sigma_2 \Delta\Psi_{l' L}+{ h.c.}, 
\end{equation}
where $C$ is the charge conjugation matrix
and $ \Psi_{l L}$ denotes the left-handed lepton
doublet with flavour $l$. 

 This interaction (\ref{eq1}) provides Majorana masses
for neutrinos   ($m_{\nu_l}= \sqrt2 h_{ll} w$).
The experimental upper bounds for the individual neutrino masses 
are 
$m_{\nu_e}\sim 2.3$ eV, $m_{\nu_\mu}\sim 170$ keV and 
$m_{\nu_\tau}\sim 18.2$ MeV \cite{numass}. Stronger yet is the bound on the sum
over neutrino masses \cite{SNO}
\begin{equation} 
0.05 < \sum_{i=e,\mu,\tau} m_{\nu_i} < 8.4~eV.
\end{equation}

The values of $h_{ll'}$ and $w$
should be consistent with these limits, implying that,
for Yukawa couplings in the range we are studying,
the value $w$ should be of the order of the Majorana mass
for left-handed neutrinos. This reinforces our assumption to neglect $w$.

Of course, the left-handed chirality of the Higgs triplet is not
something immutable. However, a reasonable theory
with natural right-handed Higgs triplet
requires extended gauge symmetry.

{\it The Left-Right Symmetric Model}, based on the gauge symmetry
$SU(2)_L\times SU(2)_R\times U(1)_{B-L}$,
 treats left-handed and
right-handed fermions symmetrically. That is,  
left-handed fermion fields transform as doublets under 
$SU(2)_L$ and as singlets under $SU(2)_R$, while the reverse
is true for right-handed fermions
\cite{LR}.
The fermionic sector contains,  in addition to SM particles,  
a right-handed neutrino for each family.
The extension of the gauge symmetry also brings new (right-handed )
gauge bosons
 $W_R$ and $Z_R$ into the model.

The scalar sector of the LR model contains many more
degrees of freedom than in the SM.
Rather than the standard Higgs doublet, it
includes a scalar bidoublet $\Phi$ with the quantum numbers 
$({2}, {2}^*,0)$; $\Phi$ gives rise to fermion Dirac masses and breaks 
the SM gauge symmetry to $U(1)_{em}$.
But, first, another Higgs field with non-vanishing $B-L$
is needed to break the 
gauge group $SU(2)_L\times SU(2)_R\times U(1)_{B-L}\,$
to the SM gauge symmetry.
If one also wants to generate  Majorana masses for the neutrino 
 through the seesaw mechanism, a triplet
 Higgs field  $\Delta_R$, with quantum numbers $(1,3,2)$  is required.
Finally, for the case of explicit $L \leftrightarrow R$
symmetry, the corresponding left-handed
triplet Higgs field should also be added:
\begin{equation}
 {\displaystyle\Delta_L=\left(\begin{array}{cc}
\Delta_L^-/ \sqrt{2}&\Delta_L^{--}\\
\Delta_L^0&-\Delta_L^- /\sqrt2
\end{array} \right)
 = (3,1,2)}
\end{equation}
For  neutrino and gauge boson masses, the presence of the 
 left-handed triplet Higgs is not  essential. However, we will
focus our attention on this representation. The $\rho$-parameter
constraints on a possible VEV for its neutral component, 
$v_L/\sqrt{2}$, hold as described above.

The most general potential describing the self-interactions
  of the scalar fields introduced above can be found in, for example
ref. \cite{MDL}.
There exist many phenomenological bounds on the parameters
  of this potential, only some of which are important here.
In particular, the mass spectrum of the scalar sector is determined 
by the
scalar potential. It is important to note
that the doubly charged Higgs triplets remain practically
unmixed \cite{GunionLR} under our assumption that the VEV of
$\Delta_{L}$ is negligible.
At the same time this assumption again leads to the relationship of
equation  (\ref{massrelation}) 
for $\Delta_L^{--}, \Delta_L^-, \Delta_L^A $,
although their masses in the LR model are no longer proportional to $v_L$.
We will use these properties of the scalar mass spectrum in our 
calculations.

The Yukawa interactions of the Higgs triplets with fermions in the model
read:

\begin{equation}
-{\cal L }_{Yuk}=i h_{R,ll'}\Psi_{lR}^TC\sigma_2\Delta_R\Psi_{l'R} + 
i h_{L,ll'}\Psi_{'L}^TC\sigma_2\Delta_L\Psi_{l'L}
\:\:\:+\:{\rm h.c.}, 
\label{Yuk}
\end{equation}
  where
$l,l'$ are flavour indices.
Along with the bidoublet's Yukawa interactions, this yields the usual 
quark $3 \times 3$ mass matrix
and charged lepton masses, while for the neutrino
one obtains a seesaw mass matrix.
Of most relevance here is the constraint that
left-handed neutrinos should be practically massless. This restricts 
the vacuum expectation value, $v_L$, of the neutral member of the 
Higgs triplet 
$\Delta_{L}$ to be small, as in our discussion of the LHTM above.
Thus, it is possible to drop the effects of
$v_L$ in our calculations.

The product of Yukawa couplings $h_{ee}h_{\mu\mu}$ dictates the 
magnitude of the process we consider here.
Existing phenomenological constraints on  
$h_{ll'}$ are as follows \cite{swartz,hm,fujii,chk}:
\begin{itemize}
\item
The rare  decays $\mu\rightarrow \bar{e} ee$ \cite{mu3e,pdb} and
$\mu\rightarrow e\gamma$ \cite{mueg}  yield  very stringent restrictions on the
non-diagonal couplings $h_{e\mu}$:
\begin{eqnarray}
h_{e\mu}h_{ee}&<&3.2\times 10^{-11}\,{\rm GeV}^{-2} M_{\Delta}^2,
\nonumber\\
h_{e\mu}h_{\mu\mu}&<&2\times 10^{-10}\,{\rm GeV}^{-2} M_{\Delta}^2.
\end{eqnarray}
Consequently, we choose to neglect all non-diagonal couplings here.
\item
From Bhabha scattering one obtains the following upper limit for $h_{ee}$
\cite{swartz},
\begin{equation}
h_{ee}^2\sim 9.7\times 10^{-6}\,{\rm GeV}^{-2} M_{\Delta}^2.
\label{hee}
\end{equation}
\item
The $(g-2)_\mu$ measurement \cite{g-2} provides an upper limit for $h_{\mu\mu}$,
\begin{equation}
h_{\mu\mu}^2\sim 2.5\times 10^{-5}\,{\rm GeV}^{-2} M_{\Delta}^2.
\label{hmm}
\end{equation}
\item
From muonium-antimuonium transition measurements 
one finds the following bound, 
which is the
most stringent at present \cite{muoniumex,swartz,fujii,chk}:
\begin{equation}
h_{ee}h_{\mu\mu}\sim 2\times 10^{-7}\,{\rm GeV}^{-2} M_{\Delta}^2.
\label{hem}
\end{equation}
The form of this bound, on the product $h_{ee}h_{\mu\mu}$, is directly
relevant to the process we consider here.
\item
For the third generation Yukawa couplings $h_{\tau\tau}$,
$h_{\tau e}$ and $h_{\tau\mu}$ there are no limits at present.
\end{itemize}

\section{Calculations and Results}

We are studying the sensitivity to
doubly charged Higgs bosons in the process $e^-\gamma \to e^+ 
\mu^-\mu^-$. The signal of 
like-sign dimuons is distinct and SM background free, 
offering excellent potential for $\Delta^{--}$ discovery.  
The Feynman diagrams describing the direct production of doubly 
charged Higgs bosons are shown in Fig. 1a. Additionally, 
the non-resonant 
contributions of the Feynman diagrams of Fig. 1b contribute to the 
distinct $\mu^-\mu^-$ signal when $M_\Delta > \sqrt{s_{e\gamma}}$.
These non-resonant contributions play an important role in
the reach that one can obtain for doubly charged Higgs masses.

To calculate the cross section we must convolute the backscattered 
laser photon spectrum, $f_{\gamma/e}(x)$, with the subprocess 
cross section, $\hat{\sigma}(e^- \gamma \to e^+ \mu^-\mu^-)$;
\begin{equation}
\sigma= \int dx  f_{\gamma/e}(x,\sqrt{s}/2) \;
\hat{\sigma}(e^- \gamma \to e^+ \mu^-\mu^-).
\end{equation}
The backscattered
photon spectrum is given in Ref. \cite{backlaser}.  Beyond a certain 
laser energy  $e^+e^-$ pairs are produced, which significantly degrades
the photon beam. This leads to a maximum $e\gamma$ centre of mass energy 
of 
$\sim 0.91 \times \sqrt{s}$.

We calculated the subprocess cross section 
with two different approaches as a 
cross-check of  
our results.  In the first we obtained analytic expressions for the 
matrix elements using the CALKUL helicity amplitude 
method \cite{CALCUL}
and performed the phase space integrals using Monte-Carlo integration
techniques.  This 
approach offers a nice check using the gauge invariance 
properties of the sum of the amplitudes.  As the expressions for the 
matrix elements  are lengthy and not particularly illuminating we do 
not include them here.  As a further check we 
compared our numerical results with those obtained using 
the COMPHEP computer package \cite{comphep}.

Because we are including contributions to the final state that proceed 
via off-shell $\Delta^{--}$'s we must include the doubly-charged Higgs 
boson width in the $\Delta^{--}$ propagator.  The $\Delta$ width, however, 
is dependent on the parameters of the model, 
which determine the size and relative importance of various
decay modes.  For example, whether  
the decays  $\Delta^{--} \to \Delta^- W^-$
and $\Delta^{--} \to \Delta^- \Delta^- $
are  allowed  depends both on the model's couplings and on the Higgs mass 
spectrum, the latter consideration determining 
whether the decays are kinematically allowed.
The decay $\Delta^{--} \to W^- W^-$ is negligible under our assumption
that the triplet's VEV is small.
The details of the model therefore can 
lead to fairly large variations in the predicted width.  
To account for this possible 
variation in widths without restricting ourselves to 
specific scenarios we calculated the width using
\begin{equation}
\Gamma (\Delta^{--}) = \Gamma_b + \Gamma_f
\end{equation}
where $\Gamma_b$ is the partial width to final state bosons and 
$\Gamma_f$ is the partial width into final state fermions.  We consider 
two scenarios for the bosonic width: a narrow width scenario with 
$\Gamma_b=1.5$~GeV and a broad width scenario with $\Gamma_b=10$~GeV. 
These choices represent a reasonable range for various values of the 
masses of the different Higgs bosons.
The partial width to final state fermions is given by
\begin{equation}
\Gamma (\Delta^{--}\to \ell^- \ell^-) = \frac{1}{8\pi} 
h^2_{ \ell \ell} M_\Delta
\end{equation}
Since we assume $h_{ ee} =h_{\mu\mu} =h_{\tau\tau} \equiv h$,
we have 
$\Gamma_f = 3 \times \Gamma (\Delta^{--}\to \ell^- \ell^-) $. Many
studies assume the $\Delta$ decay is entirely into leptons; for small
values of the Yukawa coupling and relatively low $M_{\Delta}$ this leads
to a width which is considerably more narrow than our assumptions for 
the partial width into bosons. Hence, we will also show some results for 
the case $\Gamma = \Gamma_f$, both in order to further illustrate the
width dependence and to better be able to show a connection with other 
results.

A final note before proceeding to our results is that we only consider 
chiral couplings of the $\Delta$ to leptons.  Our results here are all
based on left-handed couplings.  However,  we also did the 
calculations for right-handed couplings.  The amplitude squared and 
hence the numerical results are identical in both cases.  As a result, 
the discovery potential for $\Delta_R$ would be the same as that 
for $\Delta_L$, assuming one can make 
the same assumptions regarding parameters and mass spectra. This 
assumption may not be valid.   
We did not consider 
the case of mixed chirality.

To obtain numerical results we take as the SM inputs
$\sin^2\theta_W = 0.23124$ and $\alpha=1/128$
\cite{pdb}.  Since we work only to leading order in $|{\cal M}|^2$,
there is some arbitrariness in what to use for the above input,
in particular $\sin^2\theta_W$.

We consider two possibilities for the $\Delta^{--}$ signal.  In the 
first case we impose that all three final state particles be 
observed and identified.  
This has the advantage that the event can be fully 
reconstructed and as a check, the momentum must be balanced, at least 
in the transverse plane.  In the second case, we assume that the
positron is not observed, having been lost down the beam pipe.  
This case has the 
advantage that the cross section is enhanced due to divergences 
in the limit of massless fermions.  The disadvantage is 
that there will be some missing energy in 
the reaction so that it cannot be kinematically 
constrained which might lead to backgrounds where some particles in SM 
reactions are 
lost down the beam.  Although we expect these potential backgrounds
to be minimal, this issue needs to be studied with a realistic 
detector simulation and should be kept in mind.

To take into account detector acceptance we restrict 
the angles of the observed particles relative to the beam, 
$\theta_{\mu},\; \theta_{e^+}$, to the ranges $|\cos \theta| \leq 0.9$.
We further restrict the particle energies 
$E_{\mu}$, $E_{e^+} \geq 10$~GeV.  This 
cut is rather conservative and we have also obtained results with 
the looser cut 
$E_{\mu}$, $E_{e^+} \geq 2$~GeV. The limits 
obtained are 
quite insensitive to this variation in the value.  
We leave it to the 
experimentalists to optimize the specific value for this kinematic 
cut. We have assumed an identification efficiency for each of the
detected final state particles of $\epsilon = 0.9$.  
Finally, we note that in principle one could impose a maximum 
value on the muon energies so that the tracks are not so stiff that 
their charge cannot be determined.  Again, however, this depends on 
details of the detector and is best left 
for analysis by experimentalists 
in the context of a realistic detector simulation.

In Fig. 2 we show the cross sections as a function of $\sqrt{s}$ 
for the reaction $e^-\gamma \to 
e^+ \mu^-\mu^-$ for the two final state 
situations.  
Cross sections are shown for  $M_\Delta=400$, 800, and 1200~GeV with 
the value of the Yukawa coupling taken arbitrarily 
to be $h=0.1$.  The 
$\Delta^{--}$ 
width is taken to be $\Gamma_\Delta = 10 \hbox{ GeV } + \Gamma_f$. 
Fig. 2a shows the cross section when
all three final state particles are observed in the detector and Fig. 
2b when the positron goes down the beam.  
Below the $\Delta$ production threshold (i.e. $M_\Delta > 
\sqrt{s_{e\gamma}}$ ) the cross 
sections are rather small with a steep rise at threshold followed by a 
slow decrease with $\sqrt{s}$.  The cross section for the case when 
the positron is lost down the beam is similar in shape although a 
factor of roughly 3 larger in magnitude.

The resonance structure can be seen by plotting the invariant mass 
distributions of the final state same-sign muons.  This is shown in 
Fig. 3 for $M_\Delta = 200$, 400, and 800~GeV for $\sqrt{s}=500$~GeV 
for the broad width case where 
$\Gamma_\Delta = 10 \hbox{ GeV } + \Gamma_f$.  
For $\sqrt{s}$ above $M_\Delta$ 
threshold the $\Delta$ resonance is clearly 
seen.  Below the $\Delta$ production 
threshold, the cross section is much smaller.  Thus if $\sqrt{s}$ were 
above production threshold and if the doubly charged 
Higgs bosons had a large enough Yukawa coupling that it could be produced
in quantity, it would be possible to measure its mass and width.  
The invariant mass distribution for the narrow width case is 
virtually identical except 
that the cross section on the resonance peak is larger.  The 
distributions for the case when the positron is lost down the beam are 
similar in shape except for the low invariant mass region and the 
differential cross sections are several times larger in magnitude.

Although the cross sections below 
threshold are rather small, the expected luminosities at future 
$e^+e^-$ colliders are expected to be quite high with integrated 
luminosities over a few years equal to $\sim 500$~fb$^{-1}$.  Given 
that the signal for doubly charged Higgs bosons is so distinctive and 
SM background free, discovery would be signalled by even one event.
Because the value of the cross section for the process we consider is
rather sensitive to the $\Delta$ width, the potential for discovery 
of the $\Delta$ is likewise sensitive to this model dependent 
parameter. In Fig. 4, we show the contour for 
observing one event 
in the Yukawa coupling - doubly charged Higgs mass ($h-M_{\Delta}$) 
parameter space for the  
case of the final state with only the two muons detected with 
the positron lost down the beamline and for $\sqrt{s}=500$~GeV. 
The sensitivity to $\Gamma$ is demonstrated
by showing discovery limits for
the three cases of $\Gamma=\Gamma_f$, $1.5+\Gamma_f$, and  
$10+\Gamma_f$~GeV, where the bosonic width of the $\Delta$ has been 
varied. Relative 
to $\Gamma_b = 10$~GeV, the case of zero bosonic width has a sensitivity 
to the Yukawa coupling $h$ which is greater by a factor of about 5. It 
should be noted in comparing our results to 
those of Gregores {\it et al} \cite{gregores}, 
that they have not included the bosonic width.  
This is quite typical for 
doubly charged Higgs studies; however, as we show, the results are
rather sensitive to this parameter. Additionally, in ref. 
\cite{gregores},
they take an overall efficiency factor of 0.9 while ours is 
 a more conservative $(0.9)^2$. (Note also that the coupling of
ref. \cite{gregores}
is related to ours by $\lambda=h/\sqrt{2}$.) Taking into account this 
important width dependence and the fact that ours is 
a complete calculation
rather than an equivalent particle approximation, explains the difference
between the results. 

The general behavior of these sensitivity curves reflects the 
dependence of the cross section on $M_\Delta$.  The cross section, 
for the given kinematic cuts, starts out small and rises to a plateau 
before decreasing when $M_\Delta > \sqrt{s}$. The reduced cross 
section for small values of $M_\Delta$ arises because for small $M_\Delta$
the angular distribution is peaked near the beam direction so that 
not all the final state particles are observed.  This effect can be 
alleviated with a smaller angular cut.

The 63\%, 95\%, and 99\% probability for seeing one event corresponds to 
the average number of expected events of 1, 3 and 4.6. In 
Fig. 5, we show these three contours in the $h-M_{\Delta}$ plane 
for two cases, $\sqrt{s} = 500$~GeV and 1500~GeV. 
In each case, the results are shown for the three observed particle
final state with $\Gamma=1.5+\Gamma_f$~GeV, the narrow width case. In
the remaining figures, we present only the 95\% probability (3 event)
contours.

In Fig. 6  and 7
we show 95\% probability contours as a function of the 
Yukawa coupling and  $M_\Delta$. In each case, we assume the narrow width 
$\Gamma=1.5+\Gamma_f$~GeV case.  
Figure 6 corresponds to the center 
of mass energies 
considered for a high energy linear $e^+e^-$ collider; $\sqrt{s}=500$, 
800, 1000, and 1500~GeV.  
In Fig. 6a, the results are for the case of three observed particles in the
final state, whereas Fig. 6b shows the case where only the two muons are
observed.
Fig. 7 corresponds to the energies being 
considered for the CLIC $e^+ e^-$ collider; $\sqrt{s}=3$, 5, and 8~TeV.
Again, Fig. 7a and 7b show the results for the three body and two body
final states, respectively.
In each case, for $\sqrt{s}$ above the $\Delta$ production threshold, 
the process is sensitive to the existence of the $\Delta^{--}$ with 
relatively small Yukawa couplings.  However, when the  $M_\Delta$ 
becomes too massive to be produced the values of the Yukawa couplings 
which would allow discovery grow larger slowly.  We summarize the 
discovery potential limits for the various scenarios in Table I for 
$h =0.1$. In that Table, we present the 95\% probability mass discovery
limits for all the collider energies which we have considered, for the
case of the narrow width $\Delta$; the limits for the broad width case
for this value of $h$ are essentially the same.

\section{Summary}

Doubly charged Higgs bosons arise in one of the most straightforward 
extensions of the standard model; the introduction of Higgs triplet 
representations.  Their observation would signal physics outside the 
current paradigm and perhaps point to what lies beyond the SM.  As 
such, searches for doubly charged Higgs bosons should be part of the 
experimental program of any new high energy facility.  In this paper 
we studied the sensitivity of $e\gamma$ collisions to doubly charged 
Higgs bosons. We found that if $\sqrt{s_{e\gamma}}> M_\Delta$ doubly 
charged Higgs bosons could be discovered for even relatively small 
values of the Yukawa couplings; $h > 0.01$. For larger values of the 
Yukawa coupling the $\Delta$ should be produced in sufficient quantity 
to study it's properties.   For values of $M_\Delta$ greater than the 
production threshold, discovery is still possible for $M_\Delta$ 
greater than
$\sqrt{s}$ because of the distinctive, background free final 
state in the process  $e\gamma \to e^+ \mu^-\mu^-$ which can proceed 
via virtual contributions from intermediate $\Delta$'s.  Thus, even an 
$e^+e^-$ linear collider with modest energy has the potential to 
extend $\Delta$ search limits significantly higher than can be 
achieved at the LHC.

\acknowledgments

This research was supported in part by the Natural Sciences and Engineering 
Research Council of Canada.  N.R. is partially supported by
 RFFI Grant 01-02-17152  (Russian Fund of Fundamental Investigations).
S.G. and P.K. thank Dean Karlen and Richard Hemingway
for useful discussions.

\newpage
\begin{figure}
\centerline{\epsfig{file=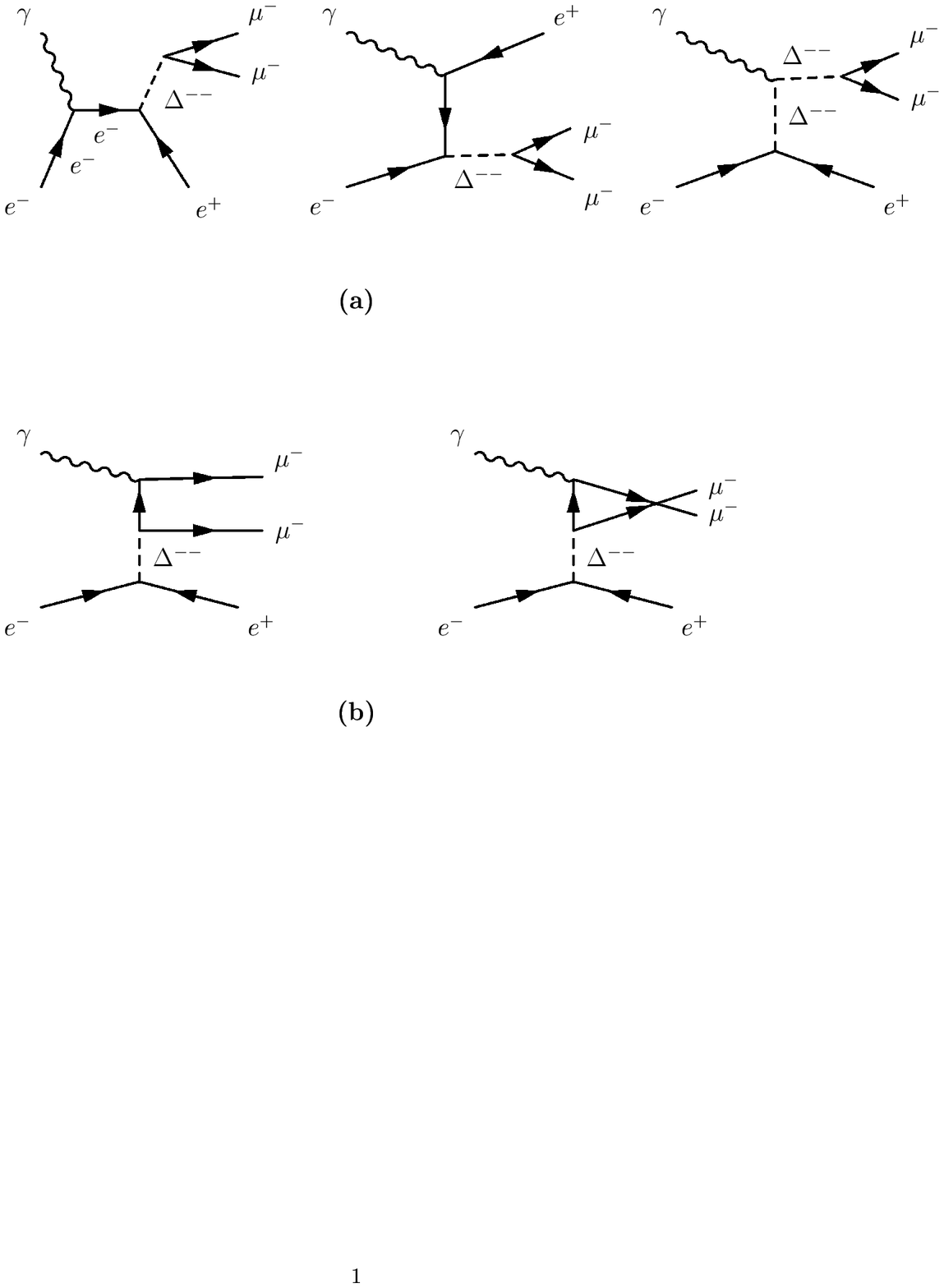,width=6.5in}}
\vspace{20pt}
\caption{The Feynman diagrams contributing to doubly charged Higgs 
boson production in $e^- \gamma \to e^+ \mu^-\mu^-$.}
\label{Fig1}
\end{figure}

\newpage
\begin{figure}
\centerline{\epsfig{file=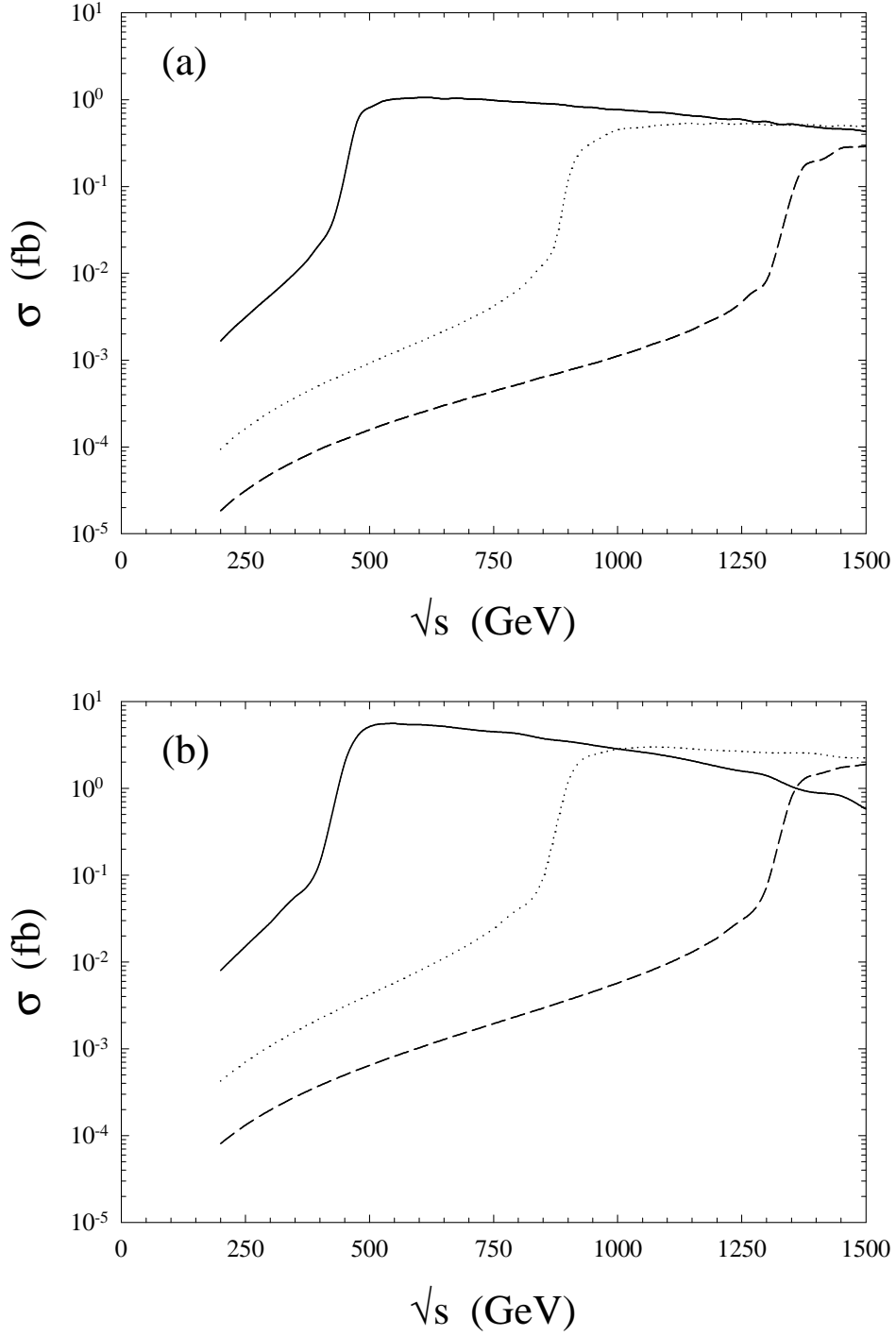,width=5.0in}}
\vspace{20pt}
\caption{The cross section, $\sigma(e^- \gamma \to e^+ \mu^-\mu^-)$
as a function of $\sqrt{s_{ee} }$. In both cases
the solid line is  for $M_\Delta = 
400$~GeV, the dotted line for $M_\Delta =800$~GeV and the dashed line 
for $M_\Delta = 1200$~GeV. (a) is for all three final state particles 
being detected and (b) is when only the $\mu^-\mu^-$ pairs are 
observed and the positron is lost down the beam pipe.}
\label{Fig2}
\end{figure}

\newpage
\begin{figure}
\centerline{\epsfig{file=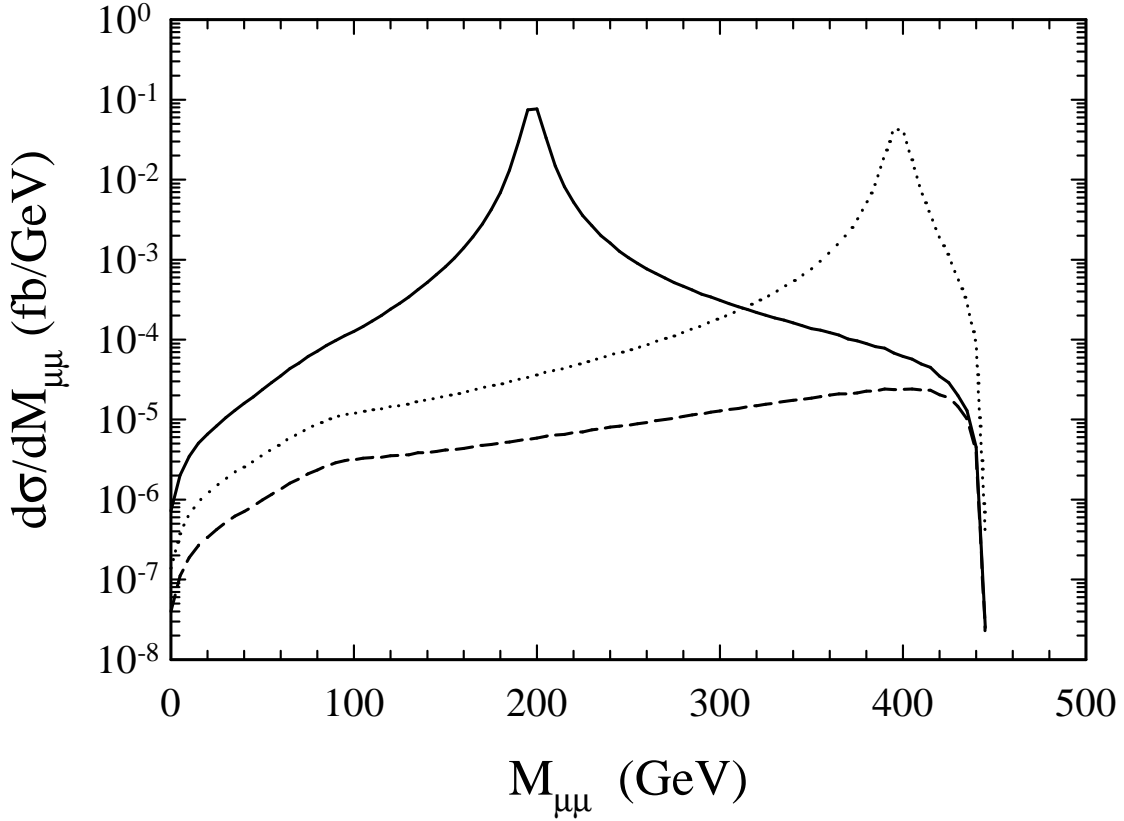,width=6.0in}}
\vspace{20pt}
\caption{The invariant mass distribution of the final state like-sign 
muons in the process $e^- \gamma \to e^+ \mu^-\mu^-$ for 
$\sqrt{s}=500$~GeV.
The solid line is  for $M_\Delta = 
400$~GeV, the dotted line for $M_\Delta =800$~GeV and the dashed line 
for $M_\Delta = 1200$~GeV. All curves are for $\Gamma_\Delta=10 + 
\Gamma_f$.}
\label{Fig3}
\end{figure}

\newpage
\begin{figure}
\centerline{
\begin{turn}{-90}
\epsfig{file=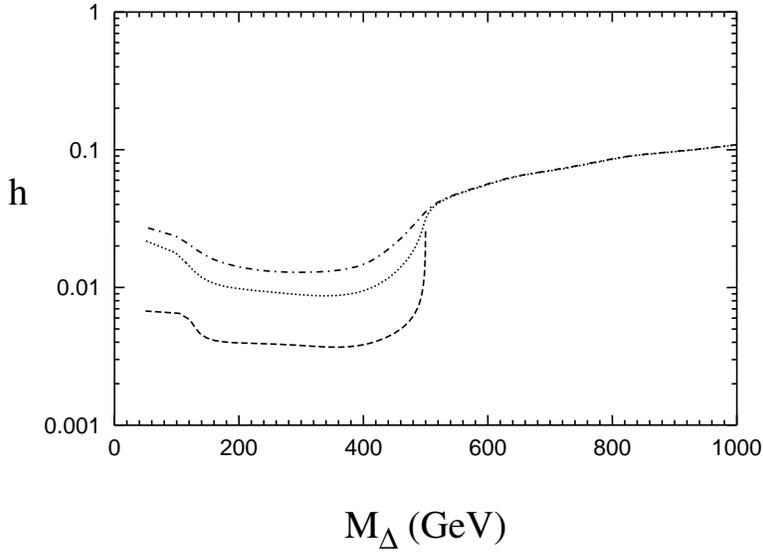,width=7.5cm,clip=}
\end{turn}
}
\vspace{20pt}
\caption{Discovery limits for doubly charged Higgs bosons as a 
function of the Yukawa coupling and $M_\Delta$ with $\sqrt{s}=500$
for different $\Gamma_\Delta = \Gamma_b + \Gamma_f$ scenarios. 
The dashed curve is for $\Gamma_b =0$, the dotted curve for 
$\Gamma_b =1.5$~GeV (the narrow width case), and the dot-dashed curve 
for $\Gamma_b =10$~GeV (the wide width case).
The limits are based on observation of one event assuming an
integrated luminosity of ${\cal L}=500$ fb$^{-1}$.
}
\label{Fig4}
\end{figure}

\newpage
\begin{figure}
\centerline{
\begin{turn}{-90}
\epsfig{file=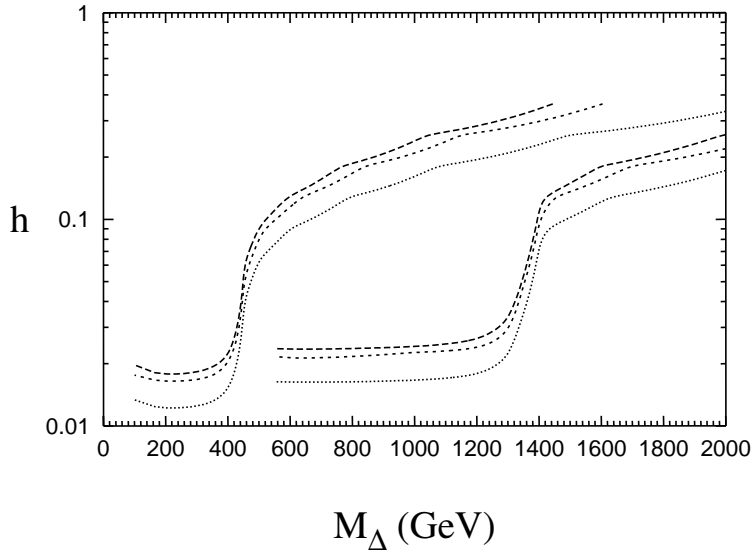,width=7.5cm,clip=}
\end{turn}
}
\vspace{20pt}
\caption{Discovery limits for doubly charged Higgs bosons as a 
function of the Yukawa coupling and $M_\Delta$ for different discovery 
probabilities.  The dotted curves are based on 63\% probability
corresponding to  
1 expected event, the dashed curves are based on 
95\% probability 
(3 events),
and the long-dashed curves are based on 99\% probability 
(4.6 events).  The 3 curves on the left 
are for $\sqrt{s}=500$~GeV and those on the right are for 
$\sqrt{s}=1500$~GeV.  In all cases
an integrated luminosity of ${\cal L}=500$ fb$^{-1}$ is assumed.
}
\label{Fig5}
\end{figure}

\newpage
\begin{figure}
\centerline{
\begin{turn}{-90}
\epsfig{file=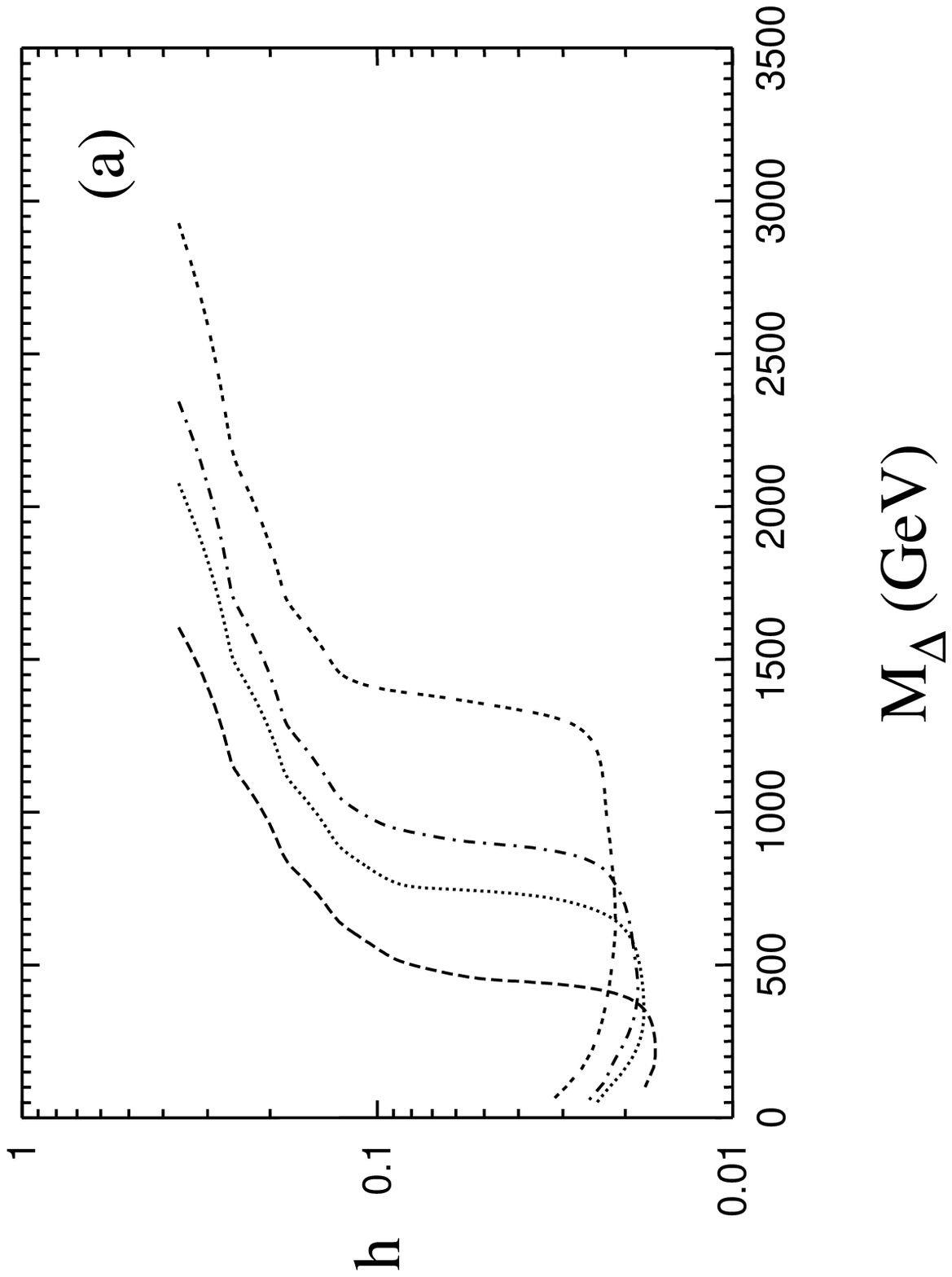,width=7.5cm,clip=}
\end{turn}
}
\centerline{
\begin{turn}{-90}
\epsfig{file=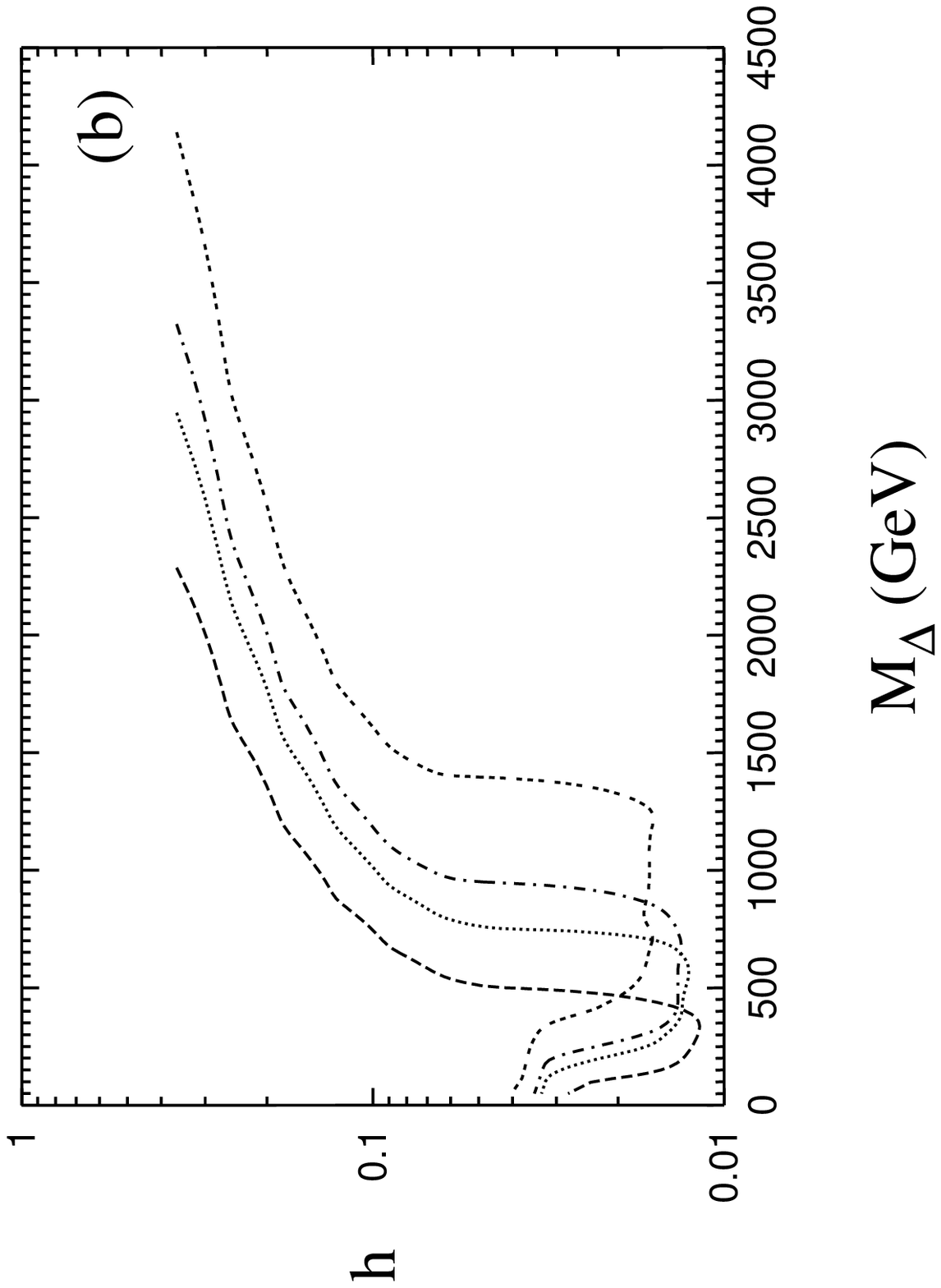,width=7.5cm,clip=}
\end{turn}
}
\vspace{20pt}
\caption{Discovery limits for doubly charged Higgs bosons as a 
function of the Yukawa coupling and $M_\Delta$ for center of mass 
energies of $\sqrt{s}=500$, 800, 1000, and 1500~GeV, 
appropriate to the LC.  The limits are based on 95\% probability of 
discovery assuming an integrated luminosity of ${\cal L}=500$ fb$^{-1}$.
(a) is for the case of all three final state particles being observed 
and (b) is the case of only the two final state muons being observed.
}
\label{Fig6}
\end{figure}

\newpage
\begin{figure}
\centerline{
\begin{turn}{-90}
\epsfig{file=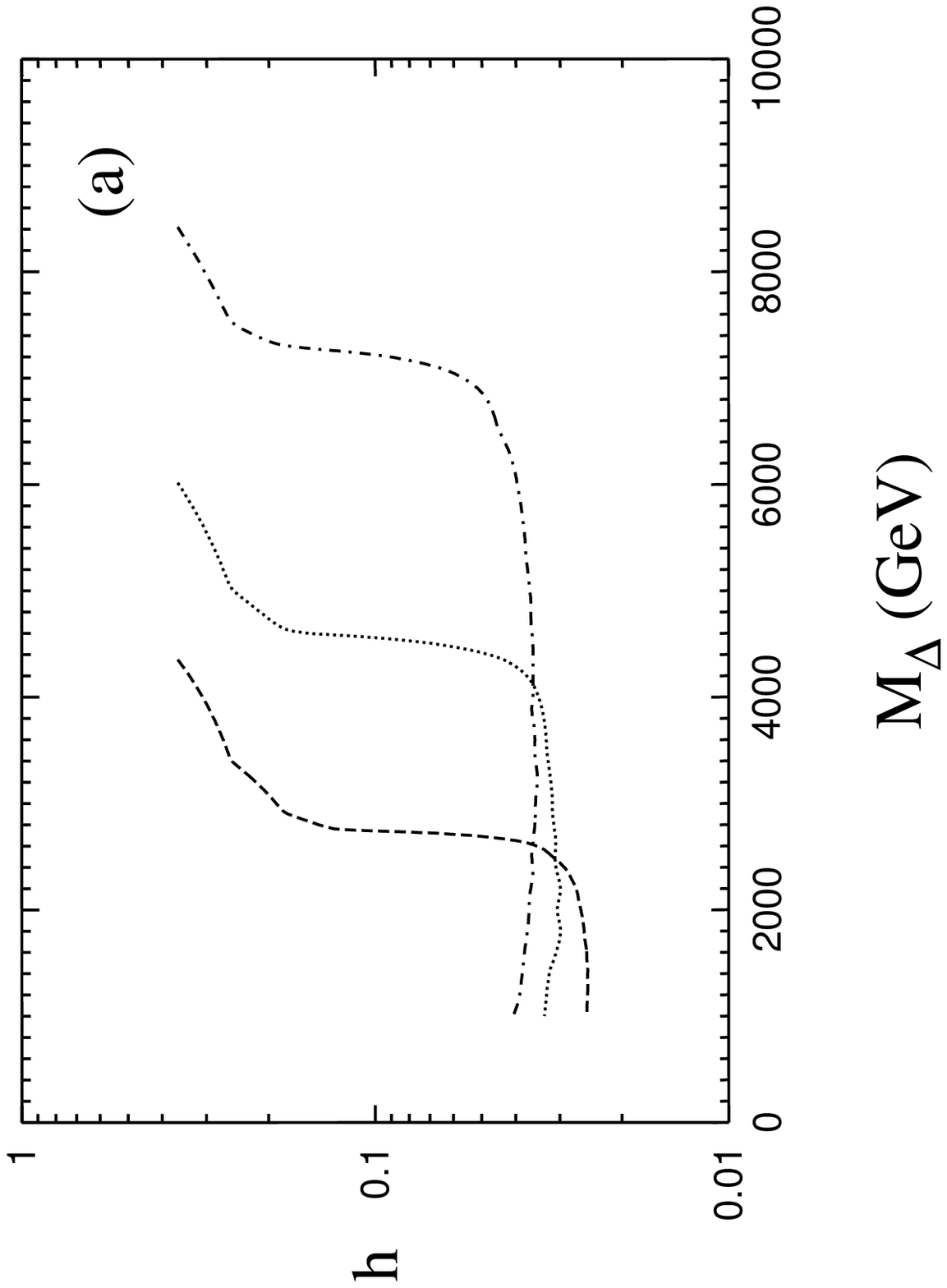,width=7.5cm,clip=}
\end{turn}
}
\centerline{
\begin{turn}{-90}
\epsfig{file=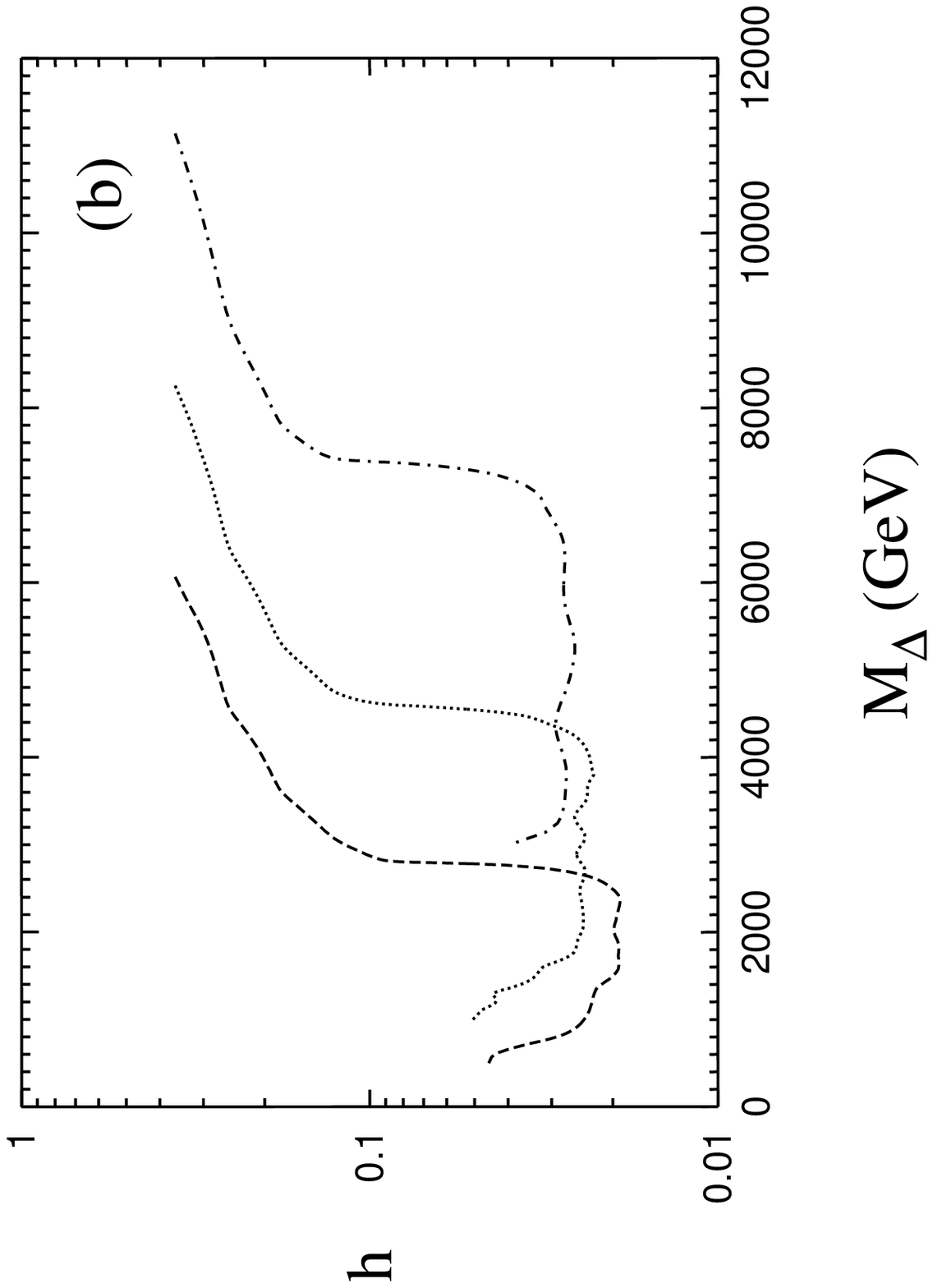,width=7.5cm,clip=}
\end{turn}
}
\vspace{20pt}
\caption{Discovery limits for doubly charged Higgs bosons as a 
function of the Yukawa coupling and $M_\Delta$ for center of mass 
energies of $\sqrt{s}=3$, 5, and 8~TeV, 
appropriate to CLIC.  The limits are based on 95\% probability of 
discovery assuming an integrated luminosity of ${\cal L}=500$ fb$^{-1}$.
(a) is for the case of all three final state particles being observed 
and (b) is the case of only the two final state muons being observed.
}
\label{Fig7}
\end{figure}

\newpage
\begin{table}[t]
\caption{95\% probability mass discovery limits of doubly charged 
Higgs bosons,
given in TeV, in $e\gamma$ collisions.  
Results are shown  for 
$\sqrt{s}=500$, 800, 1000,  1500 GeV appropriate to the  LC and 
$\sqrt{s}=3$, 5, and 8 TeV appropriate to CLIC.  In all cases we 
assume an integrated luminosity of ${\cal L}=500$ fb$^{-1}$.
The cases shown are for $e^+ \mu^-\mu^-$ detected and for the $e^+$ 
lost down the beam  for the narrow  
$\Delta$ case with the representative
Yukawa coupling of $h =0.1$. Results for the broad $\Delta$ case are
essentially the same.
}
\label{limitstabbl}
\vspace{0.4cm}
\begin{center}
\begin{tabular}{lcc}
$\sqrt{s}$ & {$e^+\mu^-\mu^-$ observed} &  
	{$\mu^-\mu^-$ observed}  \\
(TeV)      & $M_{\Delta}$ (TeV)  &
	$M_{\Delta}$ (TeV)  \\
\hline
0.5  & 	 0.54  & 0.71   \\
0.8  & 	 0.78  & 0.98   \\
1.0  & 	 0.95  & 1.15   \\
1.5  & 	 1.38  & 1.57   \\
3.0  & 	 2.72  & 2.83   \\
5.0  & 	 4.51  & 4.58   \\
8.0  & 	 7.21  & 7.30   \\
\end{tabular}
\end{center}
\end{table}


\begin{references}

\bibitem{LR} J. C. Pati and A. Salam, Phys. Rev D {\bf 10}, 275 (1974);
R. N. Mohapatra and J. C. Pati, Phys. Rev. D {\bf 11}, 566, 2558 (1975);
G. Senjanovich and R. N. Mohapatra, Phys. Rev. D {\bf 12}, 1502 (1975);
R. N. Mohapatra and R. E. Marshak, Phys. Lett. B {\bf 91}, 202 (1980);
R. N. Mohapatra and D. Sidhu, Phys. Rev. Lett. {\bf 38}, 667 (1977).

\bibitem{seesaw}
M. Gell-Mann, P. Ramond and R. Slansky, {\sl Supergravity},
ed. P. van Niewenhuizen and D. Z. Friedman
(North-Holland 1979);
T. Yanadiga, {\sl Proceedings of Workshop on Unified
Theories and Baryon Number in the Universe}, ed.
O. Sawada and A. Sugamoto (KEK, Tsukuba, 1979);
R. N. Mohapatra and G. Senjanovich, Phys. Rev. Lett.
{\bf 44}, 912 (1980).


\bibitem{Gelmini} G.B. Gelmini and M. Roncadelli, {Phys. Lett.} B
{\bf 99}, 411 (1981).

\bibitem{swartz}
M.L.~Swartz, Phys. Rev. D {\bf 40}, 1521 (1989).

\bibitem{hm}
K. Huitu and J. Maalampi, Phys. Lett. B {\bf 344}, 217 (1995).

\bibitem{fujii}
H. Fujii, Y. Mimura, K. Sasaki, and T. Sasaki, Phys. Rev. D {\bf 49}, 
559 (1994).

\bibitem{chk} 
D. Chang and W.-Y. Keung, Phys. Rev. Lett. {\bf 62}, 2583 (1989).

\bibitem{muoniumex}
L. Willmann {\it et al.}, Phys. Rev. Lett. {\bf 82}, 49 (1999).

\bibitem{framras}
P.H. Frampton and A. Rasin, Phys. Lett. B {\bf 482}, 129 (2000).

\bibitem{pleitez}
V. Pleitez, Phys. Rev. D {\bf 61}, 057903 (2000).

\bibitem{datta}
A. Datta and A. Raychaudhuri, Phys. Rev. D {\bf 62}, 055002 (2000).

\bibitem{gunion}
J.F. Gunion, C. Loomis, and K.T. Pitts, 
{\sl Proceedings of the 1996 DPF/DPB Summer Study on New 
Directions for High Energy Physics - Snowmass 96}, ed. D.G. Cassel, L. 
Trindele Gennari, and R.H. Siemann,  Snowmass, CO, 1996, p. 603
[hep-ph/9610237].

\bibitem{cuypers} 
F. Cuypers and S. Davidson, Eur. Phys. Jour. C {\bf 2}, 503 (1998);
G. Barenboim, K. Huitu, J. Maalampi,  and M. Raidal,
Phys.Lett. {\bf B394}, 132 (1997).

\bibitem{emem}
J.F. Gunion, Int. J. Mod. Phys. A {\bf 11}, 1551 (1996);
M. Raidal, Phys. Rev. D {\bf 57}, 2013 (1998);
F. Cuypers and M. Raidal, Nucl.Phys. {\bf B501}, 3 (1997).

\bibitem{london}
N. Lepor\'e, B. Thorndyke, H. Nadeau and D. London, 
Phys. Rev. D {\bf 50}, 2031 (1994).

\bibitem{gregores}
E.M. Gregores, A. Gusso, and S.F. Novaes, hep-ph/0101048 (2001).

\bibitem{rizzo}
For earlier work see T. Rizzo, Phys. Rev. D {\bf 27}, 657 (1983).

\bibitem{backlaser} 
I.F.\ Ginzburg {\it et al.}, Nucl.\ Instrum.\ Methods {\bf 205}, 47
(1983); {\it ibid} {\bf 219}, 5 (1984);
V.I. Telnov, Nucl.\ Instrum.\ Methods A {\bf 294}, 72 (1990);
C.\ Akerlof, Report No.\ UM-HE-81-59 (1981; unpublished).

\bibitem{tesla}
J.~A.~Aguilar-Saavedra {\it et al.}  [ECFA/DESY LC Physics Working Group
                  Collaboration],
hep-ph/0106315.

\bibitem{nlc}
T.~Abe {\it et al.}  [American Linear Collider Working Group Collaboration],
SLAC-R-570 (May 2001) 436p.,
hep-ex/0106058(part 1), hep-ex/0106056 (part 2), hep-ex/0106057 (part 3), 
and hep-ex/0106058
(part 4).

\bibitem{jlc}
N.~Akasaka {\it et al.}, {\sl JLC design study}, KEK-REPORT-97-1.

\bibitem{clic}
R. W. Assmann et al., CLIC Study Team, A 3-TeV
$e^+e^-$ Linear Collider Based on CLIC Technology, Ed. G. Guignard
CERN 2000-08;
M. Battaglia, CLIC Note 474 LC-PHSM-2001-072-CLIC (2001) and A. De 
Roeck private communication.

\bibitem{Godbole} R. Godbole, B. Mukhopadhyaya, and M. Nowakowski,
{Phys. Lett.} B 
{\bf 352}, 388 (1995); D.K.~Ghosh, R.M. Godbole, B. Mukhopadhyaya,
Phys. Rev. D {\bf 55}, 3150 (1997).

\bibitem{LEP} The LEP Collaborations, {Phys. Lett.} B {\bf 276}, 247 (1992).

\bibitem{numass} C. K.  Jung, in Proceedings of the EPS HEP '99
 Conference,
July 15-21, 1999, Tampere, Finland, eds. K. Huitu, H. Kurki-Suonio,
J. Maalampi (IOP publishing 2000).

\bibitem{SNO} 
Q.R.~Ahmad {\it et al.}, [The SNO Collaboration], nucl-ex/0106015.

\bibitem{MDL}
 R. Mohapatra, Phys. Rev. D {\bf 34}, 909 (1986); 
 R. Barbieri and R. Mohapatra, Phys. Rev. D {\bf 39}, 1229 (1989); 
N. G. Deshpande, J. F. Gunion, B. Kayser,
 F. Olness, Phys. Rev. D {\bf 44}, 837 (1991);
 P.Langacker and S. Uma Sankar, Phys. Rev D {\bf 40}, 1569 (1989);
J. Polak and M. Zralek, Phys. Rev. D {\bf 46}, 3871 (1992);
J. Polak and M. Zralek, Nucl. Phys. B {\bf 363}, 385 (1991);
J.Maalampi, "Signatures of Left-Right Symmmetry at High
            Energies" 
            in Proceedings of {\it The 2nd Talinn Symposium
on Neutrino Physics } eds. I.Ots and L.Palgi (Tartu 1994) p. 30.

\bibitem{GunionLR} J. F. Gunion,  J, Grifols, A. Mendez, B. Kayzer 
and F. Olness, Phys. Rev. D {\bf 40}, 1546 (1989).

\bibitem{mu3e} 
U. Bellgardt {\it et al.}, Nucl. Phys. B {\bf 299},1 (1988).

\bibitem{pdb} 
C.\ Caso {\it et al.}, Particle Data Group, 
Eur.\ Phys.\ J.\ C {\bf 3}, 1 (1998).

\bibitem{mueg}
M. L. Brooks {\it et al.}, Phys. Rev. Lett. {\bf 83}, 1521 (1999).

\bibitem{g-2}
H. N. Brown {\it et al.}, hep-ex/0102017

\bibitem{CALCUL} 
R.\ Kleiss and W.J.\ Stirling, Nucl.\ Phys.\ B {\bf 262}, 235 (1985).

\bibitem{comphep}
P. A. Baikov et al., Physical Results by means of CompHEP, in Proc. of X
Workshop on High Energy Physics and Quantum Field Theory (QFTHEP-95), eds.
B. Levtchenko, V. Savrin, Moscow, 1996, p. 101 , hep-ph/9701412;
E. E. Boos, M. N. Dubinin, V. A. Ilyin, A. E. Pukhov,
 V. I. Savrin, hep-ph/9503280.




\end{references}
\end{document}